\documentclass[final,1p,times,number]{elsarticle}

\usepackage{hyperref}
\usepackage{physics}
\usepackage{amsmath}
\usepackage{comment}
\usepackage{float}
\usepackage{units}
\usepackage{amssymb}
\usepackage{xcolor}
\usepackage{subcaption}
\usepackage{caption}
\usepackage{bm}
\usepackage{etoolbox}
\usepackage{mathtools}
\usepackage{amsfonts}

\journal{Annals of Physics}

\newcommand{\be}{ \begin{equation} }
\newcommand{\ee}{\end{equation}}
\newcommand{\bea}{ \begin{eqnarray} }
\newcommand{\eea}{\end{eqnarray}}

\newcommand{\bg}{ \begin{gather} }
\newcommand{\eg}{\end{gather}}

\renewcommand{\Im}{\mathop{\rm Im}}

\graphicspath{{figures/}}

\begin{document}

\begin{frontmatter}

\title{Tail states and unusual localization transition in low-dimensional Anderson model with power-law hopping}

\author{K.~S.~Tikhonov}
\address{Skolkovo Institute of Science and Technology, Moscow, Russia}
\address{L.D.Landau Institute for Theoretical Physics, Chernogolovka, Russia}

\author{A.~S.~Ioselevich}
\address{Condensed Matter Physics Laboratory, National Research University "Higher School of Economics", Moscow, Russia}
\address{L.D.Landau Institute for Theoretical Physics, Chernogolovka, Russia}

\author{M.~V.~Feigel'man}
\address{L.D.Landau Institute for Theoretical Physics, Chernogolovka, Russia}
\address{Skolkovo Institute of Science and Technology, Moscow, Russia}

\begin{abstract}
We study deterministic power-law  quantum hopping model with  an amplitude $J(r) \propto  - r^{-\beta}$ 
and  local Gaussian disorder  in low dimensions $d=1,2$  under the condition $d < \beta < 3d/2$.
We demonstrate unusual combination of exponentially decreasing density of  the "tail states" 
and localization-delocalization transition (as function of disorder strength $w$) 
pertinent to a small (vanishing in thermodynamic limit) fraction of eigenstates.
In a broad range of parameters density of states $\nu(E)$
decays into the tail region $E <0$ as simple exponential, $ \nu(E) = \nu_0 e^{E/E_0} $, while characteristic energy $E_0$ 
varies smoothly across edge localization transition. We develop simple analytic theory which describes
$E_0$ dependence on power-law exponent $\beta$, dimensionality $d$ and disorder strength $W$, and 
compare its predictions with  exact diagonalization results. At low energies within the bare "conduction band",
 all eigenstates are localized due to strong quantum interference at $d=1,2$; however  localization length grows fast
with energy decrease, contrary to the case of usual Schrodinger equation with local disorder.

\end{abstract}


\end{frontmatter}

\section{Introduction}

Theoretical analysis of great number of various physical problems is related to a solution of linear eigenfunction problem
of unusual type, where long-range hopping together with local disorder are crucial.
Eigenfunction problem with deterministic power-law hopping amplitudes $J(r) \propto  - r^{-\beta}$
and on-site disorder was studied in numerous papers~\cite{Levitov89-90,Burin89,Malyshev2000,Malyshev2003,Malyshev2004,Malyshev2005,Syz15,Kravtsov18,Kravtsov19,Burin20},
see also recent review~\cite{Syz-review}. In the recent paper~\cite{TF2020} two of us demonstrated that the same
mathematical framework is relevant, in particular, for a quantum superconductor-metal transition in 2D disordered metals.

In the present paper we aimed at more detailed study of eigenfunction spectrum density $\nu(E)$ and eigenfunction
statistics in such a problem; specifically, we consider below the case of low spatial dimensions $d=1,2$ and assume that
the  hopping exponent $\beta$  is in the range $d < \beta < \frac32d$, known for the most unusual behavior~\cite{Malyshev2000,Malyshev2003}. We also assume that statistics of local disorder potential is Gaussian,
with a variance $W$ of local disorder potential.

Now we review the most important known results found in Refs.~\cite{Malyshev2000,Malyshev2003,Malyshev2004,Malyshev2005,Syz15}
and demonstrate that the problem is not yet exhausted by previous papers. Crucial role of the inequality
$\beta < \frac32d$,  demonstrated in Refs.~\cite{Malyshev2000,Malyshev2003} follows from the fact that in a finite-size 
clean system of length $L$, spacing between the ground state and the first excited state scales as $\Delta_1 \propto L^{d-\beta}$ 
at large $L$. Thus at large $L$ it becomes larger than disorder matrix elements which scale as $W \cdot L^{-d/2}$.
Thus sufficiently weak disorder is unable to mix neighboring levels enough to make them localized.  Qualitatively 
similar situation is known for the nearest-neighbor hopping problem on Random Regular Graph (RRG) with local disorder~\cite{Biroli2010,Kabashima2010}, where hard gap exists between the ground state and the continuous spectrum 
(in the limit of infinite RRG, $N \to \infty$).  Upon increase of local disorder potential, ground-state wavefunction becomes localized for disorder strength $W$ above some critical value $W_c$. Details of this unusual localization transition
depend upon the type of distribution of disorder potential $P(U_i)$.

Most part of results of the papers~\cite{Malyshev2000,Malyshev2003,Malyshev2004,Malyshev2005} refers to the box distribution 
of local disorder $U(r) \in (-\frac{\Delta}{2}, \frac{\Delta}{2})$ (apart from analytic Renormalization Group study~\cite{Malyshev2003} performed for Gaussian disorder), and $W = \Delta/\sqrt{12}$.
 In this case spectral density is bound from below at $W < W_c$ in the thermodynamic limit, and the notion of the
 edge state makes sense in disordered system as well.  The paper~\cite{Malyshev2005} demonstrates 
(for any finite system size $L$)  existence of the whole narrow band of delocalized states with energies $E$
 in close proximity to the edge state energy $E_{min}$; number of these states is sub-linear with system volume $L^d$.
Straightforward generalization of the arguments present in Ref.~\cite{Malyshev2005} leads to the prediction that
 $N_{deloc} \propto L^{\gamma d}$, where  $\gamma = \frac{3d/2-\beta}{d+1 -\beta} $, which for $d=2$
reduces to universal value $\gamma_2=\frac12$ for any $\beta$.
All other eigenstates are localized, as it is expected for low-dimensional random system.
 
On the contrary, analytic analysis in the paper~\cite{Syz15} is performed for the Gaussian disorder model and 
apparently predicts localization-delocalization transition as function of $W$ and without any explicit dependence 
on the eigenstate  energy $E$. No clear distinction is made in Ref.~\cite{Syz15} between low-dimensional case
 $d=1,2$ and higher dimensions $d \geq 3$.  We believe that these results of Ref.~\cite{Syz15} may refer to 
 the last case (high dimensions) only;
at $d=1,2$ usual quantum interference (weak localization) corrections neglected in~\cite{Syz15,Malyshev2003} 
 should be strongly relevant.
It should be mentioned that the system of two RG equations differs between Ref.~\cite{Syz15} and Ref.~\cite{Malyshev2003}
although the renormalized "effective charges" are apparently the same.  In more details, while the RG equation for effective
strength of disorder coincide, the second equation - for the 
the energy renormalization factor - is written in these papers rather differently;
we cannot say at the moment which one of them is correct.

Another piece of analysis in Ref.~\cite{Syz15} is devoted to the DoS in the region of strongly localized
states which appear due to strong fluctuations of disorder potential (analogue of classical Lifshitz tails). 
Under the condition $\beta < 3d/2$, optimal fluctuation is local (determined by the lattice cut-off), similar to the case of 
usual Schrodinger equation with local disorder in high dimensions $d > 4$, see ~\cite{Harris81,Kim85,Suslov94}.
The result obtained in Eq.(3.4) of Ref.~\cite{Syz15} reads: $\nu(E<0) \propto e^{(C_0 + C_1 |E|)^2/W^2}$ with some 
(unspecified) parameters $C_{0,1}$. On the other hand, our own numerical analysis (present in Ref.~\cite{TF2020}
 and below in this paper) demonstrates purely exponential decay of $\nu(E<0) \propto e^{E/E_0}$ with $|E|$ increase in
 a wide range of parameters $\beta$ and $W$. It was shown in Ref.~\cite{TF2020} that such a pure exponential
 behavior is essential in physical applications: it leads to a power-law distributions of quantum relaxation 
rates of non-linear localized modes associated with the localized eigenstates in the DOS tail.  Therefore
we feel it  to study DOS in the tail region in more details; in particular, we are going to investigate
the dependence of characteristic energy $E_0$ on the space dimensions $d$ and exponent $\beta$,
 and to elucidate the reasons for the existence of a purely exponential behavior in a broad range of $\nu(E)$ variation.

Important consequence of the tail states in presence of Gaussian-distributed disorder
is the absence of a well-defined edge of the spectrum.  Rare localized states appear at $E <0$ due to disorder fluctuations 
even at sub-critical disorder strength $W < W_c$, contrary to the case of box distribution  studied in papers~\cite{Malyshev2004,Malyshev2005}.  However, as we will show below, 
strongly localized eigenstates of the same origin are present at some narrow range of positive energies $E>0$,
 where they co-exist with delocalized eigenstates discussed above.  Co-existence of very different kinds of
 eigenstates in the same energy range is one of our main results of the present paper.

The rest of the paper is organized as follows.  Sec.2 is devoted to formulation of the model. Sec.3 contains
self-consistent theory of the effects of typical (weak) disorder. Sec.4 is devoted to the effects of strong 
local fluctuations which lead to the tail states, here we present both analytical and numerical (for $d=1$) results.
In Sec.5 we analyze unusual localization transition near the spectrum edge
 as function of disorder strength $W$, and present 1D exact-diagonalization results in  for the distribution function
of participation ratios $\mathcal{P}(P_2)$ which demonstrate co-existence of localized and extended states in the same
energy stripe.
 Sec. 6 contains analysis of the "conduction band" states corresponding to moderately low positive $E$ in the
two-dimensional version of our model;  we find energy dependence of localization length $L(E)$ and show that it grows
with $E$ decrease. In Sec. 7 we present numerical data for localization length in 1-dimensional case at moderately
low energies; the same phenomenon of $L(E)$ growth with energy decrease is demonstrated.
Sec. 8 contains our conclusions.

\section{The model}

\begin{figure}[tbp]
\includegraphics[width=0.48\textwidth]{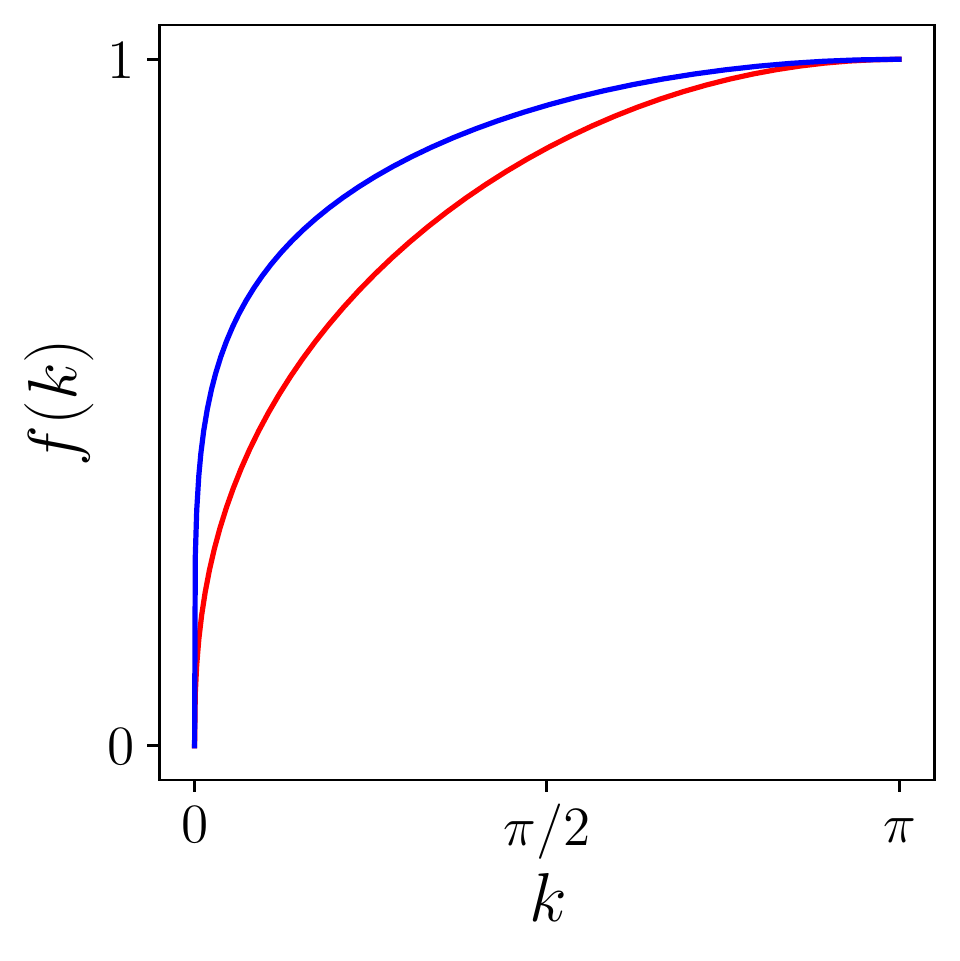}
\includegraphics[width=0.48\textwidth]{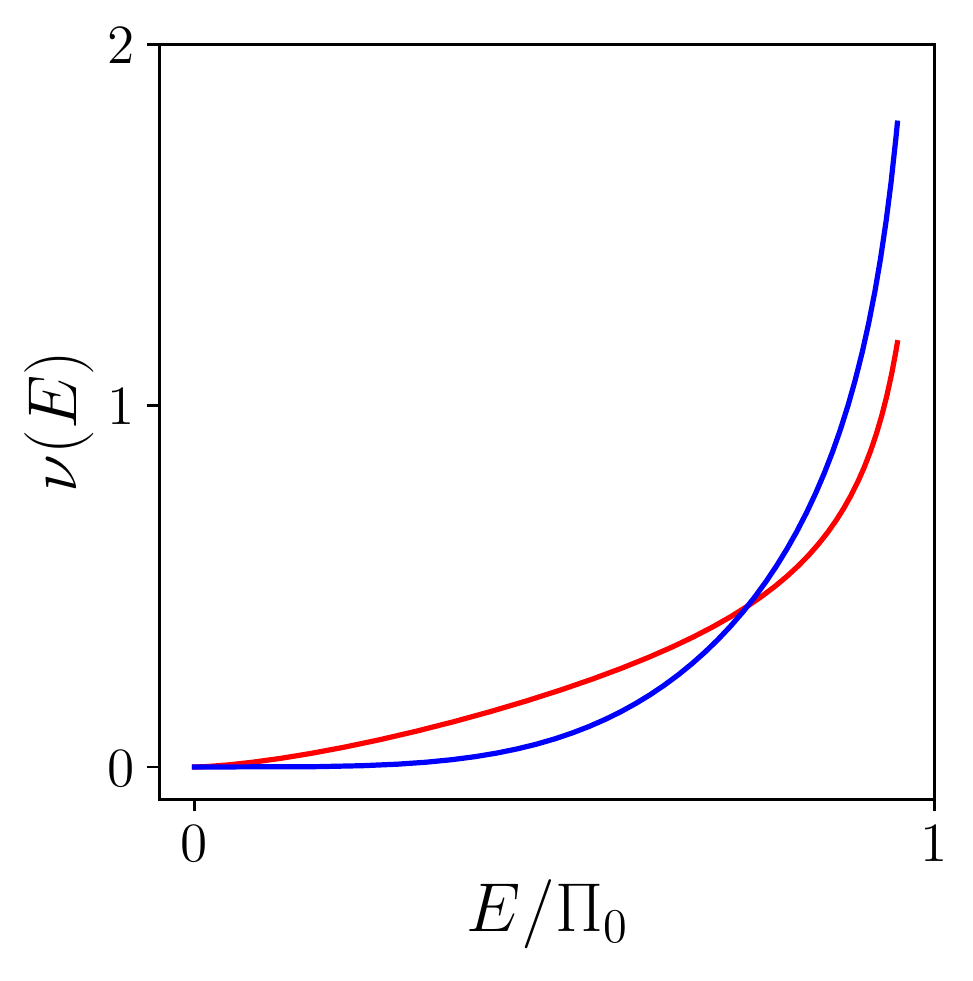}
\caption{Spectrum (left), see Eq. (\ref{fq2}) and DOS (right) for 1D system with $\alpha=0.4$ (red) and $\alpha=0.2$ (blue).}
\label{fig:fnu}
\end{figure}

We consider Anderson model with long-range hopping and local Gaussian disorder.  The Hamiltonian is
\begin{equation}
H = \sum_{\mathbf{r},\mathbf{r}'} \Pi_{\mathbf{r}-\mathbf{r}'} a^+_\mathbf{r} a_\mathbf{r'} +
\sum_\mathbf{r} V_\mathbf{r} a^+_\mathbf{r} a_\mathbf{r}
\label{H1}
\end{equation}
where coordinate vectors $\mathbf{r}$, $\mathbf{r}'$ run over 1D or 2D lattice, the coordinates of the lattice sites belong to
the range  $(-\frac{L}{2},\frac{L}{2})$. Lattice  dimensionality will be denoted below as $d$.
Kinetic energy operator is defined via its Fourier-transform:
\begin{equation}
 \Pi_{\mathbf{r}} = \int \frac{d^dq}{(2\pi)^d}\Pi(\mathbf{q})\exp[i \mathbf{q} \mathbf{r}]
\label{Pi1}
\end{equation}
where
\begin{equation}
\Pi(\mathbf{q}) = \Pi_0 f(q), \qquad f(q)\approx\left\{\begin{aligned} f_0q^{\alpha},& \quad(q\ll 1),\\
1,& \quad(|q|\to \pi)\end{aligned}\right.
\label{Pi2}
\end{equation}
The corresponding density of states $\nu_0(E)$ in the absence of disorder
\begin{eqnarray}
\nu_0(E)=\int \frac{d^dq}{(2\pi)^d}\delta (E - \Pi(\mathbf{q})),\\
\nu_0(E) \approx  \frac{C}{\Pi_0}\left(\frac{E}{\Pi_0}\right)^{\frac{d}{\alpha}-1}\theta(E),\quad E\ll \Pi_0
\label{rho0}
\end{eqnarray}
where $C= (\pi d \alpha f_0^{d/\alpha})^{-1}$.
Simplest  form of the function $f(\mathbf{q})$ for $d$-dimensional cubic lattice is
\begin{equation}
f(\mathbf{q}) =  \left(\frac1{d}\sum_{\mu=1}^d \sin^2\frac{q_\mu}{2}\right)^{\alpha/2}
\label{fq2}
\end{equation}
with $f_0= (4d)^{-\frac{\alpha}{2}}$. We will use below the form (\ref{fq2}). The functions $f(q)$ and $\nu_0(E)$ are plotted in Fig. \ref{fig:fnu}.

The quantity $\Pi_0$ determines overall bandwidth in our model.  It serves as the natural scale of energy
within "flat band" approximation formulated below and valid in the limit $\alpha \ll 1$.
For general values of $\alpha$ scaling of energies with $\Pi_0$ is valid approximately.
Below we will derive various results analytically, and compare them with numerical data obtained via 
exact diagonalization. For the latter, we use energy units corresponding to the following choice of the energy scale:
\begin{equation}
\Pi_0 = 2^{1+\alpha} d^{\alpha/2}
\label{Pi00}
\end{equation}
With the choice (\ref{Pi00}) low-momentum asymptotic of the spectrum is 
$\Pi(q) \approx 2|\mathbf{q}|^\alpha $, in agreement with notations used in Ref.~\citep{TF2020}.

The exponent $\alpha$ is related to the coordinate-space exponent $\beta$ as $\alpha = \beta - d$, and we are interested
in the  "high dimension" case, $ 0 < \alpha < \frac{d}{2} $ (for usual Schroedinger equation it would be realized at $d > 4$, see 
Ref.\cite{Harris81,Kim85,Suslov94}).

 Boundary conditions are periodic, and there are $L^d$ distinct sites in the lattice.
The uncorrelated on-site random potential obey the identical Gaussian distributions:
\begin{eqnarray}
{\cal P}(\{V\})\equiv\prod_iP(V_i),\quad P(V)=\frac{1}{\sqrt{2\pi}W}e^{-\frac{V^2}{2W^2}}
\label{P0}
\end{eqnarray}
We are interested in the average density of states (DOS)
\begin{eqnarray}
\langle\nu(E)\rangle=\frac{1}{\pi L^d}\sum_{\mathbf{r}}\int{\cal D}V{\cal P}(\{V\})
{\rm Im}\,G^{R}_E(\mathbf{r}|\mathbf{r})
\label{rho1}
\end{eqnarray}
where the retarded Green function $G^{R}_E(\mathbf{r}|\mathbf{r}')$ obeys the Schroedinger equation
\begin{eqnarray}
(E+i\gamma-V_\mathbf{r})G(\mathbf{r}|\mathbf{r}')-
\sum_{\mathbf{r}_1}\Pi_{\mathbf{r}-\mathbf{r}_1}G(\mathbf{r}_1|\mathbf{r}')= \delta_{\mathbf{r},\mathbf{r}'}
\label{Green1}
\end{eqnarray}
with $\gamma\to +0$.  


There are two major types of  disorder potential fluctuations: typical fluctuations
with $V_\mathbf{r} \sim W$ in all lattice sites and anomalously strong \textit{local} fluctuations with 
$V_\mathbf{r} \sim -\Pi_0$  at  very rare sites (note,  that $\Pi_0 \gg W)$. Contrary to the usual Lifshitz tails in dimensions $d < 4$,
the second type of fluctuations is crucial to obtain the eigenstates at negative (renormalized) energies.

\textbf{Flat band approximation.}
Note that for small $\alpha$ the function $f(\mathbf{q})$ is  close to 1
in the major part of the Brillouin zone, except narrow region around $|q| =0$. In the zero-order approximation, it is possible
to ignore this narrow region, in other words --  to neglect the difference between $f(\mathbf{q})$ and unity, while calculating some integrals over $d^dq$;  in what follows we will call it
"flat band approximation". As usual, "flat band" means space locality, i.e. the neglect of velocity  $\mathbf{v}=d\Pi(\mathbf{q})/d\mathbf{q}$. The flat band approximation should not be taken too seriously: particles with very small momenta have non-zero and quite high velocities, which diverge in the $q\to 0$ limit as $v(q) \propto q^{\alpha-1}$. However, the small-$q$ region of the momentum space contains a small fraction of the whole phase volume; this is why flat band approximation 
is reasonable for the integrals over momentum space.

\section{Typical fluctuations of random potential and Density of States}

The major effect of typical fluctuations with $V_\mathbf{r} \sim W$  upon the DoS can be described within
self-consistent Born approximation (SCBA), see Sec.II of Ref.~\cite{TF2020}. The corresponding equation reads~\cite{comment2}:
\begin{equation}
\nu_{scba}(E) = -\frac{1}{\pi}\Im \mathcal{G}(E + \Sigma(E))
\label{DoS-scba}
\end{equation}
where the self-energy $\Sigma(E)$ should be found from the system of nonlinear equations
\begin{eqnarray}
\label{sigma}
\Sigma(E) = - \langle V^2\rangle \mathcal{G}(E + \Sigma(E)); \\
\langle V^2\rangle=W^2,\quad \mathcal{G}(\varepsilon) = \int \frac{d^dq}{(2\pi)^d}g(\varepsilon,\mathbf{q})
\label{Sigma}
\end{eqnarray}
where
\begin{equation}
g(E,\mathbf{k}) \equiv [E - \Pi(\mathbf{k})]^{-1}
\label{G2q}
\end{equation}
is the Green function in the absence of disorder. 

The solution of Eqs.(\ref{sigma},\ref{Sigma}) can be easily obtained within the flat band approximation, where  $\mathcal{G}(\varepsilon) = \mathcal{G}_0(\varepsilon) = (\varepsilon - \Pi_0)^{-1}$ so that the equation \eqref{Sigma} is reduced to a quadratic one:
\begin{eqnarray}
\Sigma_0(E) = -\frac{W^2}{E + \Sigma_0(E)-\Pi_0}
\label{Sigma1}
\end{eqnarray}
Since we are interested in the case $|E|,W\ll\Pi_0$, the relevant solution of \eqref{Sigma1} reads
\begin{eqnarray}
\Sigma_0 \approx W^2/\Pi_0
\label{Sigma2}
\end{eqnarray}
is almost $E$-independent. It means that in the zero flat-band approximation the effect of random potential is reduced to the shift of all the energies by the same value.

It is convenient to introduce the shifted energy
\begin{eqnarray}
\varepsilon=E+\Sigma_0 
\label{Sigma2z}
\end{eqnarray}
In what follows we will use $\varepsilon$ as the energy argument while discussing major part of the spectrum, i.e. 
"conduction band".

\begin{figure}[tbp]
\centerline{\includegraphics[width=0.75\textwidth]{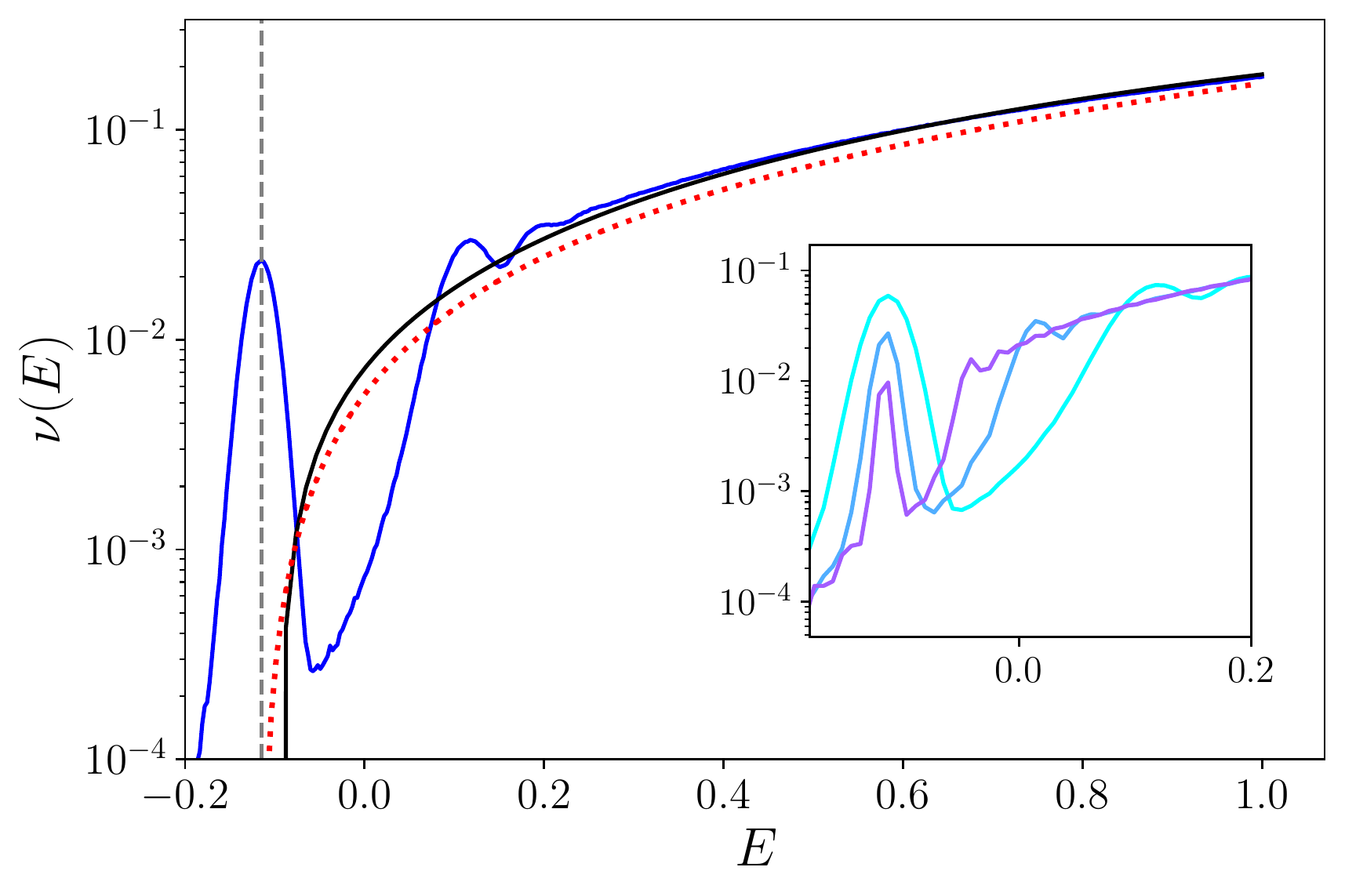}}
\caption{DOS for the $d=1$ system with disorder $W=0.45$ and $\alpha=0.4$. Blue: ED result obtained in a finite system with $L=1024$ sites. Red dashed line: $\nu_0(E-\delta)$, with $\delta=-0.115$ . Black line: DOS derived in the SCBA approximation, see Eqs. (\ref{DoS-scba}), (\ref{sigma}) and (\ref{Sigma}). Grey dashed line: position of a shifted band edge $E=\delta$. Inset: evolution of the DoS with the system size: $L=1024,\;4096,\;16384$ from cyan to magenta.}
\label{fig:scba}
\end{figure}

One can  go beyond the zero flat-band approximation and write $\Sigma(\varepsilon)=\Sigma_0+\Sigma_1(\varepsilon)$, then the self consistency equation for $\Sigma_1(\varepsilon)$ takes the form
\begin{eqnarray}
\Sigma_1(\varepsilon)\approx -\int\frac{d^dq}{(2\pi)^d}\left[1-\frac{\Pi(q)}{\Pi_0}\right]\frac{W^2}{\varepsilon+\Sigma_1(\varepsilon)-\Pi(q)},
\label{scba7}
\end{eqnarray}
The factor in the square brackets on the right hand side of \eqref{scba7} suppresses the contribution of the flat part of the band, so that the integral over $q$ here is dominated be a narrow vicinity of $q=0$. As a result,    $\Sigma_1$ is purely imaginary and can be evaluated within the Born approximation:
\begin{eqnarray}
\Sigma_1(\varepsilon)\approx \frac{i}{2\tau_1(\varepsilon)},\quad \frac{1}{\tau_1(\varepsilon)}=2\pi W^2\nu_0(\varepsilon), 
\label{scba8}
\end{eqnarray}
The condition of applicability for the Born approximation reads
\begin{eqnarray}
\varepsilon\tau_1(\varepsilon)\sim\left(\frac{\Pi_0}{W}\right)^2\left(\frac{\Pi_0}{\varepsilon}\right)^{\frac{d}{\alpha}-2}\gg 1,
\label{scba91}
\end{eqnarray}
which is fulfilled for all $\varepsilon$ provided that $\alpha<d/2$ and $W,|\varepsilon|\ll \Pi_0$. Thus, we conclude that
\begin{eqnarray}
\Sigma(\varepsilon) \approx W^2/\Pi_0+i\pi W^2\nu_0(\varepsilon)
\label{Sigma2w}
\end{eqnarray}
for all relevant energies $\varepsilon$.

We have shown that, as far as only typical fluctuations of potential are taken into account, the semi-classical approximation remains valid for all the states in the shifted conduction band and the scattering rate can be described within the standard Born approximation; actually no self-consistency condition was required.

The localization (unavoidable at $d=1,2$)  only occurs due to quantum interference effects at large distances $\sim L_{\rm loc}(\varepsilon)$. According to Eq.(\ref{scba91}), the localization length  \textit{increases} as $\varepsilon$ approaches 
the bottom of the band, which is quite unusual. 
As we will see in the following sections, such a paradoxical behavior is destroyed, if the anomalous (strong) fluctuations are taken into account. These fluctuations provide an exponentially small but finite contribution to the density of states at $\varepsilon\to 0$ and thus lead to the breakdown of the quasi-classical condition \eqref{scba91} at certain domain of exponentially small energies $\varepsilon$. In particular, it revives the need for the self-consistency equation and leads to a rapid decrease of $L_{\rm loc}(\varepsilon)$ for smallest $\varepsilon$.

In the Fig. \ref{fig:scba}, we show the DoS as found by exact diagonalization (ED) of the model Eq. (\ref{H1}) in 1D for $\alpha=0.4,\;W=0.45$ (we recall that the spectrum was normalized according to Eq.(\ref{Pi00})). The solid blue line shows the DoS at the system $L=1024$. These data are  substantially influenced by finite-size effects -- finite-size quantization peaks (the lowest few are still well resolved, although broadened by disorder). On the inset, the DOS for several system sizes is shown in a smaller energy window, from which the drift of the energy of the first excited state with the system size is obvious, as well as decrease of the weight of the lowest peak. At the same time, in a large energy window ($E\gtrsim 0.2$) the DOS at $L=1024$ has already reached  its $L\to\infty$ limit. In this range, it is instructive to compare it with the the results obtained by SCBA approximation. It is realized via numerical solution of  Eqs. (\ref{DoS-scba}), (\ref{sigma}) and (\ref{Sigma}), the results are shown by a black line. The SCBA approximation describes the $E\gtrsim 0.2$ part of the spectrum very well. Moreover, since the position of the renormalized band edge is not sensitive to finite-size effects, the black curve tends to zero exactly at the energy  of the lowest peak. Finally, the red dashed line shows a simpler approximation $\nu_0(E-\delta)$, with a numerically determined shift $\delta=-0.115$ which is slightly different from the one derived in the flat--band approximation, see Eq. (\ref{Sigma2}) which gives $\delta=-\Sigma_0\approx - 0.078$, the difference can be attributed to the fact that $\alpha=0.4$ is not sufficiently small.

\section{Anomalous tail in the density of states: strong local fluctuations of random potential}

\subsection{Local strong fluctuations}

Now we take into account strong single-site fluctuations, assuming that $V_\mathbf{r} = - V\delta_{0,\mathbf{r}}$ and $V\sim\Pi_0\gg W$.  Although the probability of such rare fluctuations is exponentially small, we will see that they are very important not only for $\varepsilon<0$, where they give rise to the tail of the density of states, but also in a certain  narrow strip of positive $\varepsilon$. Within this strip the exponentially small correction $1/2\tau_2(\varepsilon)$ to $\Im \Sigma$, arising due to single-site fluctuations, dominates over the
contribution $1/2\tau_1(\varepsilon)$ of typical fluctuations described by the expression \eqref{scba8}.
It is also important to note that strong single-site fluctuations only affect the imaginary part of self energy, the real part is always dominated by the typical ones.

For clarity, in this Subsection we consider only strong local fluctuation at certain site, totally neglecting typical fluctuations at other sites. The interplay of both types of fluctuations will be be discussed in the subsequent Subsections.

Equation for the corresponding retarded Green function $G$ in the momentum representation reads:
\begin{equation}
(E_+ -\Pi(\mathbf{k}'))G(\mathbf{k}'|\mathbf{k}) + V\int\frac{d^dq}{(2\pi)^d}G(\mathbf{q}|\mathbf{k})=
(2\pi)^d\delta(\mathbf{k}-\mathbf{k}'),
\label{Green1}
\end{equation}
where $E_+= E+ i0$ and integration goes over the Brillouin zone. 
Eq.(\ref{Green1}) can be solved explicitly:
\begin{equation}
G(\mathbf{k},\mathbf{k}')= (2\pi)^d \delta(\mathbf{k}-\mathbf{k}')g(\mathbf{k}) + \Lambda(E,V)g(\mathbf{k})g(\mathbf{k}')
\label{G2}
\end{equation}
\begin{eqnarray}
 \Lambda(E,V) = [V^{-1}(E) - V^{-1} + i0]^{-1},
\label{U}
\end{eqnarray}
where the function $V(E)$ (which is generally a complex one) is defined by
\begin{eqnarray}
 \frac1{V(E)} = -\mathcal{G}(E)=
 \frac1{V_0(E)}+ i\pi\nu_0(E),\\
 \frac1{V_0(E)}={\tt{v.p.}}\int \frac{d^dq}{(2\pi)^d}\frac1{\Pi(\mathbf{q}) - E}. 
\label{VE}
\end{eqnarray}

For $E<0$ the density of states $\nu_0(E)\equiv 0$ and  $V(E)\equiv V_0(E)$ is purely real, being
defined by the principle-value integral. In particular, within the flat band approximation we obtain
\begin{equation}
V_0(E) \approx \Pi_0  - E.
\label{VE2}
\end{equation}

At $E>0$, where  $\nu_0(E)\neq 0$, $V(E)$ acquires an imaginary part, while its real part changes continuously in the vicinity of $E=0$
Note that $\Lambda(E,V)$ is nothing else but the exact scattering amplitude. At small $V\to 0$ the perturbative Born result $\Lambda(E,V)\to -V$ is recovered, while at large $V=V(E)$ the scattering amplitude has a pole. This pole lies in the complex plane of $V$, its imaginary part is infinitesimal at $E<0$.

Naturally, a pole in the scattering amplitude corresponds to a bound state (or quasi-bound state if $V(E)$ contains nonzero
imaginary part). The wave function for this bound state can  be found easily:
\begin{eqnarray}
\psi_0(\mathbf{k})=\left[-\frac{d\mathcal{G}(E)}{dE}\right]^{-1/2}g(\mathbf{k}),
\label{ta77}
\end{eqnarray}
or, in the $\mathbf{r}$-reperesentation
\begin{eqnarray}
\psi_0(\mathbf{r})=\left[-\frac{d\mathcal{G}(E)}{dE}\right]^{-1/2}\mathcal{G}(E)\,\delta_{\mathbf{r},\mathbf{0}}+\Delta\psi(\mathbf{r}),\quad \Delta\psi(\mathbf{r})=\left[-\frac{d\mathcal{G}(E)}{dE}\right]^{-1/2}\int\frac{d\mathbf{k}}{(2\pi)^d}h(\mathbf{k},E)e^{-i(\mathbf{r}\cdot\mathbf{k})},
\label{ta77a}
\end{eqnarray}
where we have introduced
\begin{eqnarray}
h(\mathbf{k},E)\equiv g(\mathbf{k},E)-\mathcal{G}(E),\quad \int\frac{d\mathbf{k}}{(2\pi)^d}h(\mathbf{k},E)=0.
\label{ta7bs}
\end{eqnarray}
In particular, from \eqref{ta7bs} it follows that  $\Delta\psi(0)=0$ and the characteristic spatial scale for $\Delta\psi({\bf r})$ is $\sim (\Pi_0/|E|)^{1/\alpha}\gg 1$.

 Within the flat band approximation the coefficient in front of $\delta_{{\bf r, 0}}$ is close to unity and  $\Delta\psi({\bf r})$ is relatively small  for all $r$: the wave function is strongly localized at the site of the fluctuation.   However, though being small in magnitude, the tail $\Delta\psi({\bf r})$ is  decaying only as a power of $r$, therefore it may lead to interesting effects due to weak, but long-range hybridization of states. These effects will be discussed in a separate publication, here we will simply neglect $\Delta\psi({\bf r})$. This is an important simplification, since it allows to neglect completely  the overlap of wave-functions at different fluctuations and, thus, to treat their contributions as independent ones.

The sought correction $\Delta\nu(\varepsilon)$ to the DoS comes from the second term in Eq.(\ref{G2}) upon calculating the Green function trace
$\mathcal{G}(E)$ and averaging over potential distribution $P(V)$:
\begin{equation}
\Delta\nu(E) = 
\frac{1}{\pi }\int P(V)dV\,\,{\rm Im}\,\left[\frac{\frac{d}{dE}\left(\frac{1}{V(E)}\right)}{V^{-1}-V^{-1}(E)+i0}\right]
\label{DoS1}
\end{equation}
While deriving Eq.(\ref{DoS1}) we  used Eqs.(\ref{G2},\ref{U}) and the identity  
$\int \frac{d^dq}{(2\pi)^d} g^{2}(E,\mathbf{q}) \equiv -\frac{d}{dE}\int \frac{d^dq}{(2\pi)^d} g(E,\mathbf{q})$.

The contribution to the density of states due to deep and narrow fluctuations comes from the part of the integral \eqref{DoS1} over $V$, corresponding to the pole in the integrand at $V=V(E)$:
\begin{equation}
\Delta\nu(E) = 
{\rm Re}\,\left[P(V(E))\frac{dV(E)}{dE}\right]
\label{DoS11}
\end{equation}
This general result is valid for both signs of $E$.

\subsection{Interplay of rare strong local fluctuations and  typical weak fluctuations}

\begin{figure}[ht]
\centerline{\includegraphics[width=0.6\linewidth]{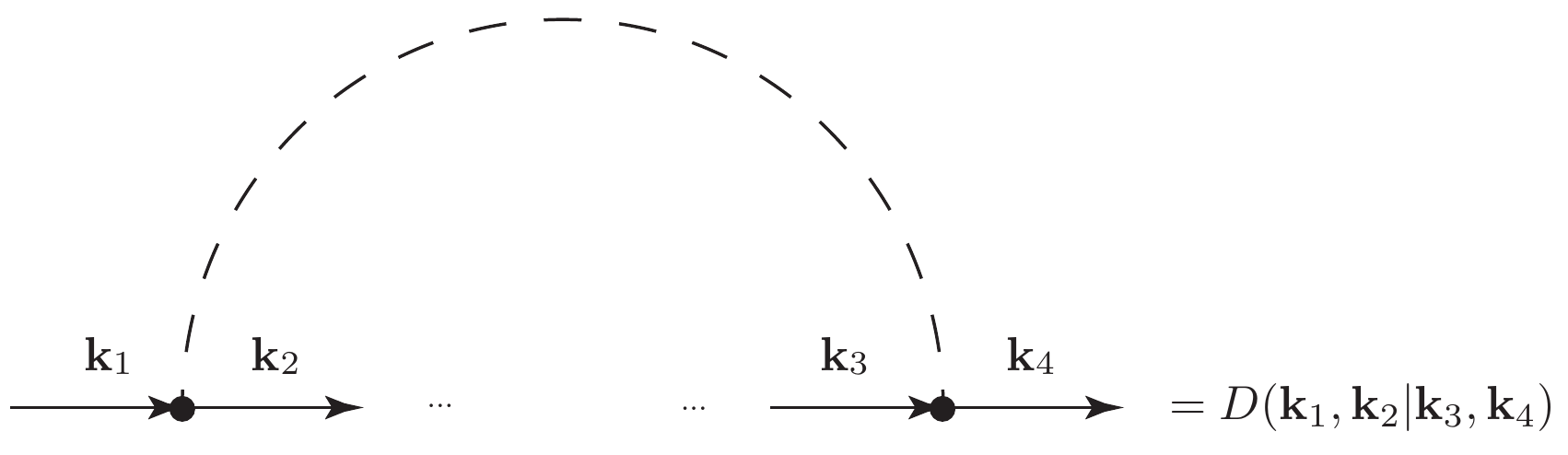}}
\caption{Diagrammatic representation of the object defined in Eq.\eqref{ta7qw3}.}
\label{diagram}
\end{figure}

Let us now take into account typical fluctuations of $V_i$ that occur at sites, other then $i=0$. It can be done in a standard diagrammatic approach, with one important modification, however. Namely, to avoid double counting, we should exclude the site $i=0$, where the value of $V$ is fixed, from the set of scatterers. 
As a result, the self-consisted Born approximation  equations with respect to weak fluctuations acquires now the form
\begin{eqnarray}
\Sigma (\mathbf{k},\mathbf{k}')=\int\frac{dk_1dk_2}{(2\pi)^2}\tilde{D}(\mathbf{k},\mathbf{k}_1|\mathbf{k}_2,\mathbf{k}')G(\mathbf{k}_1|\mathbf{k}_2),\label{se10}\\
G(\mathbf{k}_1|\mathbf{k}_2)=\left[\{\hat{G}_0^{-1}+\hat{\Sigma}\}^{-1}\right]_{\mathbf{k}_1,\mathbf{k}_2}
\label{se1}
\end{eqnarray}
 with the following change of the definition of the "impurity lines"  $D(\mathbf{k}_1,\mathbf{k}_2|\mathbf{k}_3,\mathbf{k}_4)$ 
 entering the impurity diagram technique.
Namely, for a dashed line shown in Fig. \ref{diagram}, instead of the standard definition
\begin{eqnarray}
\quad  D(\mathbf{k}_1,\mathbf{k}_2|\mathbf{k}_3,\mathbf{k}_4)=W^2(2\pi)^d \delta(\mathbf{k}_1-\mathbf{k}_2+\mathbf{k}_3-\mathbf{k}_4),\label{ta7qw0}
\end{eqnarray}
one should use a modified one:
\begin{eqnarray}
\tilde{D}(\mathbf{k}_1,\mathbf{k}_2|\mathbf{k}_3,\mathbf{k}_4)=W^2[(2\pi)^d \delta(\mathbf{k}_1-\mathbf{k}_2+\mathbf{k}_3-\mathbf{k}_4)-1],\label{ta7qw3}
\end{eqnarray}
"-1 term" in the square brackets just takes care of the site $\mathbf{r}=0$ to be excluded. Note that this term does not depend on the momenta due to point-like character of the scatterer.
In the perturbative limit one can write $G=G_0$ on the right-hand side of \eqref{se10}, so the perturbative solution is
\begin{eqnarray}
\Sigma (\mathbf{k},\mathbf{k}')=\int\frac{dk_1dk_2}{(2\pi)^{2d}}\tilde{D}(\mathbf{k},\mathbf{k}_1|\mathbf{k}_2,\mathbf{k}')G_0(\mathbf{k}_1|\mathbf{k}_2),
\label{se2}
\end{eqnarray}
and
\begin{eqnarray}
G_0^{-1}(\mathbf{k},\mathbf{k}')+\Sigma (\mathbf{k},\mathbf{k}')=(2\pi)^{2d} \delta(\mathbf{k}-\mathbf{k}')(E+\Sigma_0-\Pi(\mathbf{k}))+V-\Sigma_0,\label{se5a}\quad
\Sigma_0\equiv W^2\mathcal{G}(E),
\end{eqnarray}
In the $\mathbf{r}$-representation we get the following Schrodinger equation for the Green function $G(\mathbf{r}|\mathbf{r}')$:
\begin{eqnarray}
\left\{E+\Sigma_0-\Pi(-i\partial _{\mathbf{r}}))-U(\mathbf{r})\right\}G(\mathbf{r}|\mathbf{r}')=\delta(\mathbf{r}-\mathbf{r}'),\quad U(\mathbf{r})=-\tilde{V}\delta({\bf r})\label{se5aa}
\end{eqnarray}
We note that the expression \eqref{se5a} has the same structure, as $G_0^{-1}(\mathbf{k},\mathbf{k}')$ up to the following re-definitions
$E\to \varepsilon\equiv E+\Sigma_0,\; V\to \tilde{V}\equiv V-\Sigma_0$
so that
\begin{eqnarray}
G(\mathbf{k},\mathbf{k}')=G_0(\mathbf{k},\mathbf{k}')|_{E\to \varepsilon,\; V\to \tilde{V}}, 
\label{se6}
\end{eqnarray}
\begin{eqnarray}
\Delta\nu(E) = 
\frac{1}{\pi }\int P(V)dV\,\,{\rm Im}\,\left[\frac{\frac{d}{d\varepsilon}\left(\frac{1}{V(\varepsilon)}\right)}{\tilde{V}^{-1}-V^{-1}(\varepsilon)+i0}\right],
\label{DoS11}
\end{eqnarray}
The main contribution to the integral in \eqref{DoS11} comes from the pole of the integrand at $\tilde{V}=V(\varepsilon)$. According to definition of $\tilde{V}$, it corresponds to $V=V_{\rm eff}(\varepsilon)$, where
\begin{eqnarray}
V_{\rm eff}(\varepsilon)=V(\varepsilon)+\Sigma_0\approx \Pi_0-\varepsilon+\Sigma_0=\Pi_0-E\equiv V(E).
\label{DoS11}
\end{eqnarray}
Then, performing the integration, we get
\begin{eqnarray}
\nu(\varepsilon)=\nu_0(\varepsilon)+{\rm Re}\,\left\{P\left[V_{\rm eff}(\varepsilon)\right]\frac{dV(\varepsilon)}{d\varepsilon}\right\},
\label{DoS11a}
\end{eqnarray}

Thus, the role of weak fluctuations is reduced to two effects:
\begin{enumerate}
\item The bottom of the band is shifted by $\Sigma_0$.
\item  The bottom of the potential well is shifted by the same value $\Sigma_0$, so that the relative depth of the well $V_{\rm eff}$ is not affected.
\end{enumerate}

The simple result \eqref{DoS11a} was obtained in the leading flat-band approximation. In principle, it is not very difficult to find also a leading correction: it results in an additional (small, but long range) potential
\begin{eqnarray}
 U(\mathbf{r})\to -\tilde{V}\delta({\bf r})+\Delta U({\bf r}),\quad \Delta U({\bf r})=-\Lambda(E,V)W^2\left|\int\frac{dq}{(2\pi)^d}h(\mathbf{q})e^{i(\mathbf{q}\cdot\mathbf{r})}\right|^2\label{se5aa}
\end{eqnarray}
This additional potential leads only to small corrections to the density of states and we will not discuss them here.

\begin{figure}[tbp]
\centerline{\includegraphics[width=0.75\textwidth]{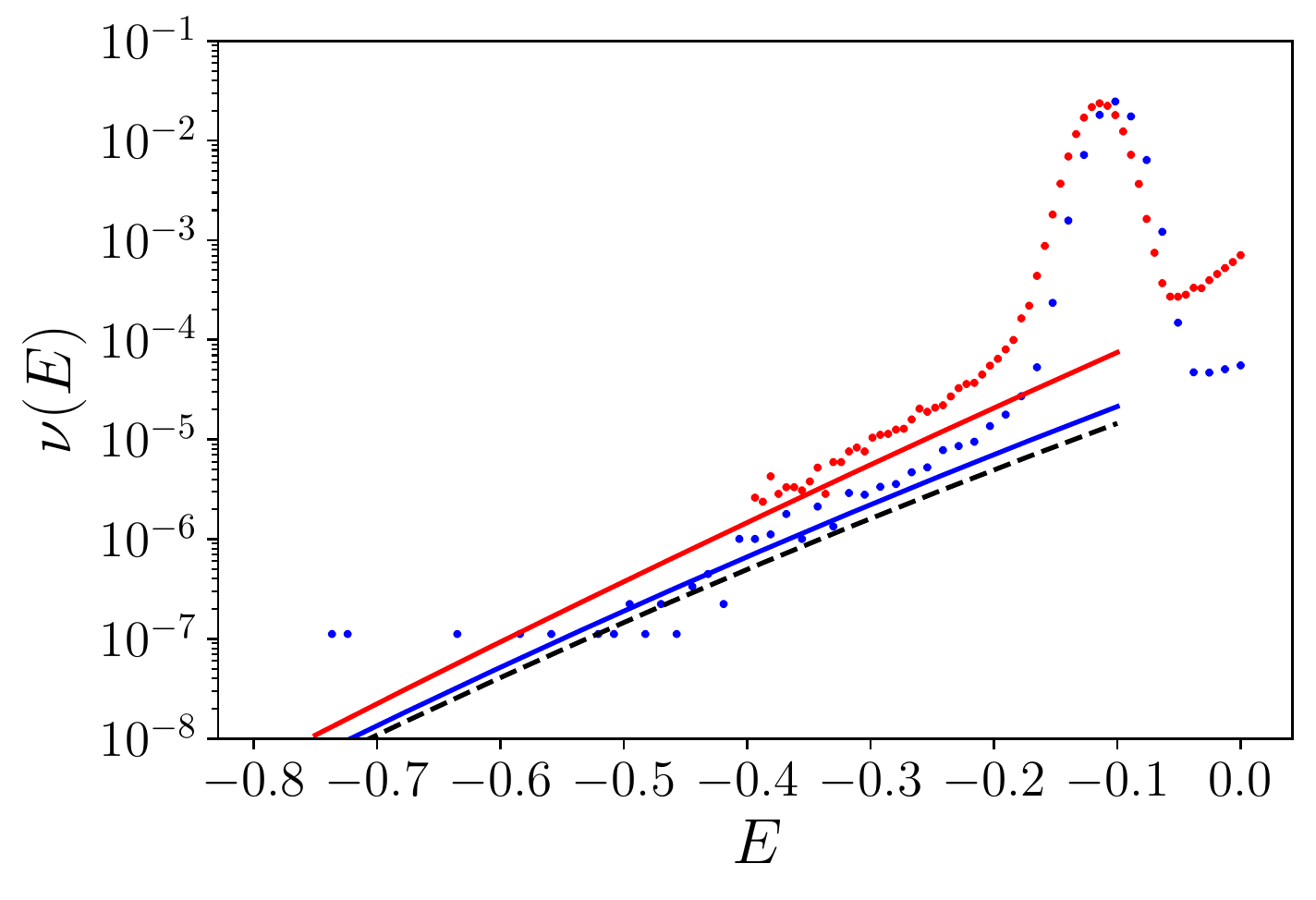}}
\caption{DOS in the tail for the $d=1$ system with disorder $W=0.45$ and $\alpha=0.4$ (red), $\alpha=0.2$ (blue). Dots: exact diagonalization result obtained in a finite system with $L=1024$ sites. Solid line: tail DoS, according to Eq. (\ref{DoS2}). Black dashed line: result of Eq. (\ref{rho-lin}) with $\alpha\to 0$.}
\label{fig:dos1d}
\end{figure}

\subsection{Negative energies: tail states}


 At negative energies renormalized $\varepsilon < 0$, the potential  $V(\varepsilon)$ is real, since bare DOS is absent, 
and we get:
\begin{equation}
\nu(\varepsilon<0)  =  \frac{1}{\sqrt{2\pi}W}\left|\frac{dV_0(\varepsilon)}{d\varepsilon}\right|\exp\left\{-\frac{V_0^2(E)}{2W^2}\right\}.
\label{DoS2}
\end{equation}
Within the flat-band approximation (for small $\alpha$), this result can be further simplified:
\begin{equation}
\nu(\varepsilon<0)  = \frac{1}{\sqrt{2\pi}W}e^{-(\Pi_0 + |E|)^2/2W^2},\quad E=\varepsilon-W^2/\Pi_0.
\label{rho-neg}
\end{equation}
At small $W$ the density of states rapidly decays with $|E|$, so that only a relatively narrow domain of $|E|\ll \Pi_0$ is observable in practice.  Then the $E-$dependence of $\nu(\varepsilon)$ is close to simple exponential in a broad range
of the DoS variation:
\begin{equation}
\nu(\varepsilon<0)  \approx \nu_0e^{-\Pi_0 |E|/W^2},\quad \nu_0=\frac{e^{-\Pi_0^2/2W^2}}{\sqrt{2\pi}W}
\label{rho-lin}
\end{equation}

In the Fig. \ref{fig:dos1d}, we show the results of ED for the density of states in the tail for $\alpha=0.4$ and $\alpha=0.2$ with $W=0.45$. The solid lines show the result of computation according to Eq. (\ref{DoS2}) and the black solid line corresponds to Eq. (\ref{rho-lin}) at $\alpha\to 0$. 

\subsection{Positive energies: resonances}

At positive (with respect to the renormalized band edge) energies DOS is non-zero even in the absence of special strongly localized eigenstates 
related to local potential fluctuations. However, the latter states also exist for $\varepsilon >0$ in the form of resonances, where they 
also contribute to the full DOS.  This contribution is essential only in a narrow domain of small positive energies
\begin{eqnarray}
\varepsilon\lesssim \varepsilon_{\mathrm{cross}}\ll\Pi_0
\label{DoS3k}
\end{eqnarray}
where $\varepsilon_{\mathrm{cross}}$ is exponentially small. As a consequence, in the above domain
\begin{eqnarray}
\Pi_0\nu_0(\varepsilon)\ll 1,\quad V(E)\approx V_0(E)\approx \Pi_0.
\label{DoS3r}
\end{eqnarray}
Then, expanding the exponent in \eqref{DoS11} in small $\nu$ we obtain 
\begin{eqnarray}
\Delta\nu(\varepsilon>0)\approx  \frac{1}{\sqrt{2\pi}W}\left|\frac{dV_0(\varepsilon)}{d\varepsilon}\right|\exp\left\{-\frac{(\Pi_0-E)^2}{2W^2}\right\}\left(1-\frac{\pi^2\nu_0^2(\varepsilon)\Pi_0^2}{2}\left(\frac{\Pi_0}{W}\right)^4\right).
\label{DoS3}
\end{eqnarray}
The second term in the brackets is the correction arising due to the existence of small but finite $\nu_0(\varepsilon)$. It  is small under the condition
\begin{eqnarray}
\Pi_0\nu_0(\varepsilon)\ll W^2/\Pi_0^2\ll 1
\label{DoS3re}
\end{eqnarray}
which is stronger than \eqref{DoS3r}. However, as we will see soon, the strong condition \eqref{DoS3re} is also likely to be fulfilled for all relevant energies $\varepsilon\lesssim \varepsilon_{\mathrm{cross}}$. 
Indeed, within the flat-band approximation we have
\begin{equation}
\Delta\nu(\varepsilon\lesssim \varepsilon_{\mathrm{cross}}) \approx \Delta\nu_0=\frac{1}{\sqrt{2\pi}W}e^{-\Pi_0^2/2W^2}
\label{DoS4}
\end{equation}
Now we are prepared to estimate the  crossover energy $\varepsilon_{\mathrm{cross}}$.
Since the contribution (\ref{DoS4}) is essential as long as $\Delta\nu_0\gtrsim\nu_0(\varepsilon)$, see Eq.(\ref{rho0}), 
 $\varepsilon_{\mathrm{cross}}$ is given by
\begin{equation}
\varepsilon_{\mathrm{cross}} \sim
\Pi_0
\left(\frac{\Pi_0}{W }\right)^{\alpha/(d-\alpha)}\exp\left\{-\frac{\Pi_0^2}{2W^2}\frac{\alpha}{d-\alpha}\right\}
\label{Ecross}
\end{equation}
Note that in the exponent in Eq.(\ref{Ecross}) small $\alpha$ in numerator can partially compensate small $W^2$
in denominator. At  higher energies $\varepsilon \geq \varepsilon_{\mathrm{cross}}$
density of states is dominated by the "conduction band" contribution (\ref{rho0}).

It is important to note, that the self-consistent Born approximation seems to work also in the range $\varepsilon \lesssim\varepsilon_{\mathrm{cross}}$. Indeed, here the strong local fluctuations dominate the density of final states for the scattering process  within the conduction band. At the same time the typical distance between these fluctuations is smaller than the wave-length of band excitations, so that the procedure of averaging of the self energy remains reasonably justified and we expect that formula
\begin{equation}
\frac{1}{\tau_{\rm out}}=\frac{1}{\tau_1(\varepsilon)}+\frac{1}{\tau_2(\varepsilon)}=W^2[\nu_0(\varepsilon)+\Delta\nu_0]
\label{tau-out}
\end{equation}
is applicable and expression \eqref{Sigma2w} should be replaced by
\begin{eqnarray}
\Sigma(\varepsilon) \approx W^2/\Pi_0+i\pi W^2[\nu_0(\varepsilon)+\Delta\nu_0].
\label{Sigma2wm}
\end{eqnarray}
As a result, the Born approximation is expected to break down at  $\varepsilon\sim W^2\nu_0$,
while at lower energies  strong localization regime is expected to be established.

The energy domain $0 < \varepsilon < \varepsilon_{\mathrm{cross}}$ is interesting since it contains eigenstates of essentially different
origin; we expect that they also have very different properties in terms of localization length. 
This combination of eigenstates could be illustrated by inspecting the distribution function of the inverse 
participation ratio $I_2 = \sum_i\psi_i^4$ characterizing eigenstates $\psi_i$ in the relevant energy range.
However, to analyze $I_2$ distribution relevant in the thermodynamic limit, we would need to perform exact 
diagonalization on extremely large systems (which is beyond the scope of the present paper);
the point is that weak dispersion (\ref{Pi2}) makes the finite-size effects rather strong even for our largest samples with 
$L=16384$, as indicated by the position of the second peak in the DOS, see Fig.~\ref{fig:dos1d}.
Instead, we concentrate in the next Section on the properties the eigenfunctions which belong to 
the left-most peak in DOS.

\section{Special eigenfunctions  near the band edge}

\begin{figure}[tbp]
\centerline{\includegraphics[width=0.75\textwidth]{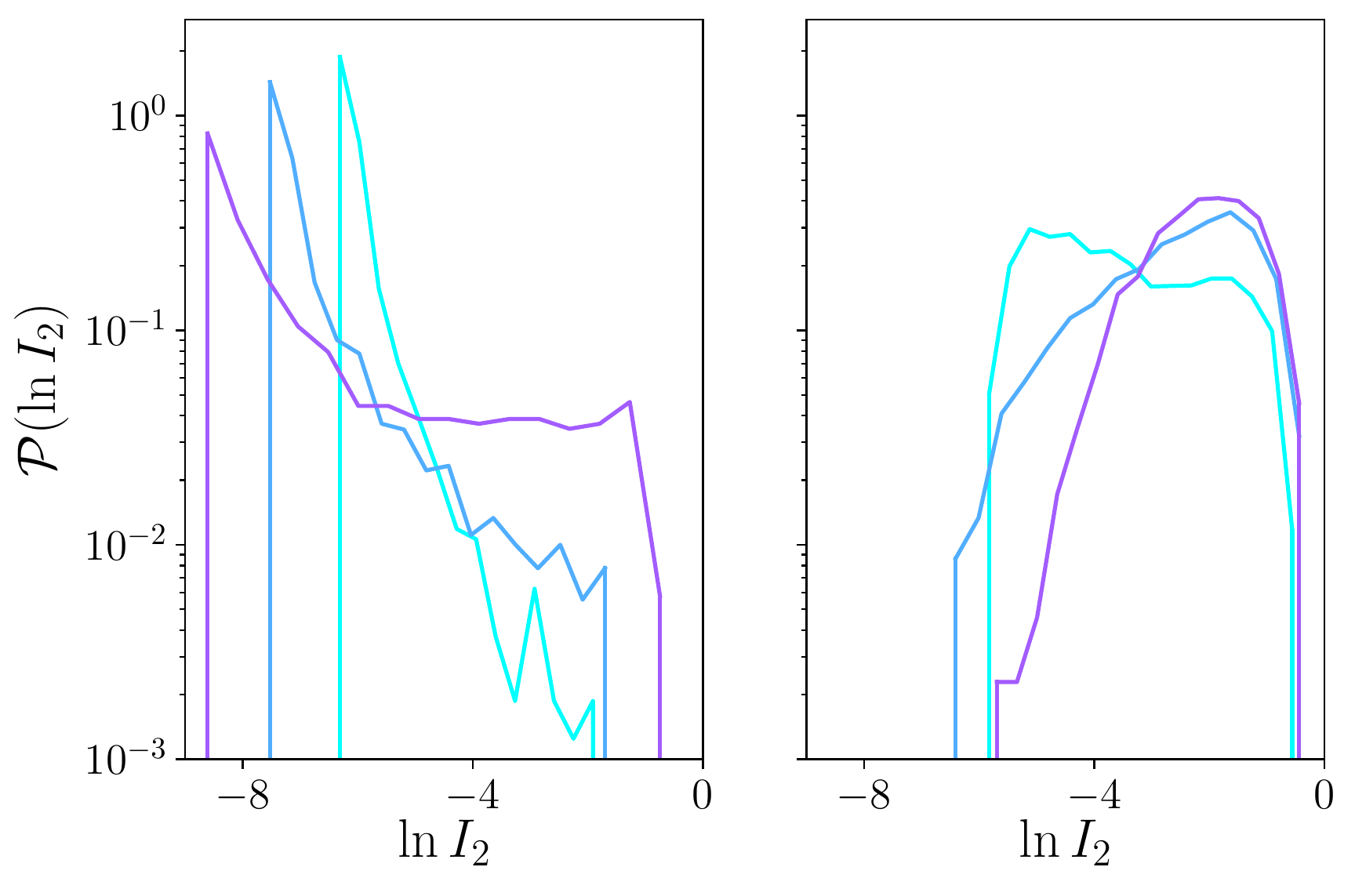}}
\caption{Distribution function of the logarithm of inverse participation ratio $\ln I_2$ of the states near the shifted band edge for $\alpha=0.4$, $W=0.45$ (left) and $W=0.63$ (right)system sizes $L= 1024, 4096, 16384$ from cyan to magenta. The function $\mathcal{P}(\ln I_2)$ is essentially zero outside the domain, marked by vertical segments on the plot.}
\label{fig:ipr1d_peak_2}
\end{figure}

In a finite large disordered system,  the eigenstates - "descendants" of the uniform $q=0$ state
 of the pure system, contribute to a leftmost peak in the DOS shown in Fig.~\ref{fig:scba}. 
They are expected~\cite{Malyshev2005} to be extended for subcritical disorder $W < W_c$ and 
localized at $W > W_c$.  We will now study distribution function of the inverse participation ratios in the relevant energy range, found by exact diagonalization on $10^4-10^3$ samples of the hopping problem (for system sizes $2^{10}-2^{14}$, correspondingly). We first show in Fig.~\ref{fig:ipr1d_peak_2}, left panel, the data for $W=0.45$ in a narrow range
of energies around the peak (for $W=0.45$, we consider $-0.135<E<-0.10$; for $W=0.63$, the corresponding range is $-0.28 < E < -0.22$).

Clearly, the increase of $L$ leads to continuous shift of the left edge of the distribution 
$\mathcal{P}(I_2)$, indicating the presence of delocalized eigenstates. On the other hand,
the right part of this distribution contains a lot of small-size eigenstates, which co-exists - in the same
narrow range of $E$ - with delocalized ones. Moreover, the total share of the localized eigenstates grows
with $L$ increase, according to Ref.~~\cite{Malyshev2005}. We conclude that $W_c > 0.45$ for $\alpha=0.4$.
Distribution function $\mathcal{P}(I_2)$ corresponding to the largest system size $L=16384$ demonstrates clear
bi-modal behavior (recall that it is plotted in logarithmic scale in Fig.~\ref{fig:ipr1d_peak_2}).
We conjecture that small-size eigenstates originate from local large fluctuations discussed in Sec.IV B.
Co-existence of qualitatively different types of eigenstates in the same narrow energy stripe is unusual,
but not exceptional:  such a feature was recently found in a Quantum Random-Energy Model~\cite{FFI2019}.

Right panel of Fig.~\ref{fig:ipr1d_peak_2} shows data of the same kind for larger disorder $W=0.63$.
Now evolution of the distributions with the system size is completely different. Namely, the position of the 
left edge of the distribution
$\mathcal{P}(I_2)$ is non-monotonic with the size.  
It is shifted a little bit to the left while $L$ grows from 1024 to 4096 (although this shift is twice  smaller than it was
for $W=0.45$), while for largest $L=16384$ the left edge is located at the largest $I_2$ value among all three sizes.
These data indicate that maximal localization length for $W=0.63$ is somewhat larger than 4000; the left edge of 
distribution for the largest $L=16384$ is supposed to saturate at the position corresponding to $L=4096$ upon increase
of statistics. We conclude that the critical value of disorder $W_c < 0.63$.


\section{The case of $d=2$: weak localization at moderately low positive energies}

\begin{figure}[tbp]
\centerline{\includegraphics[width=0.75\textwidth]{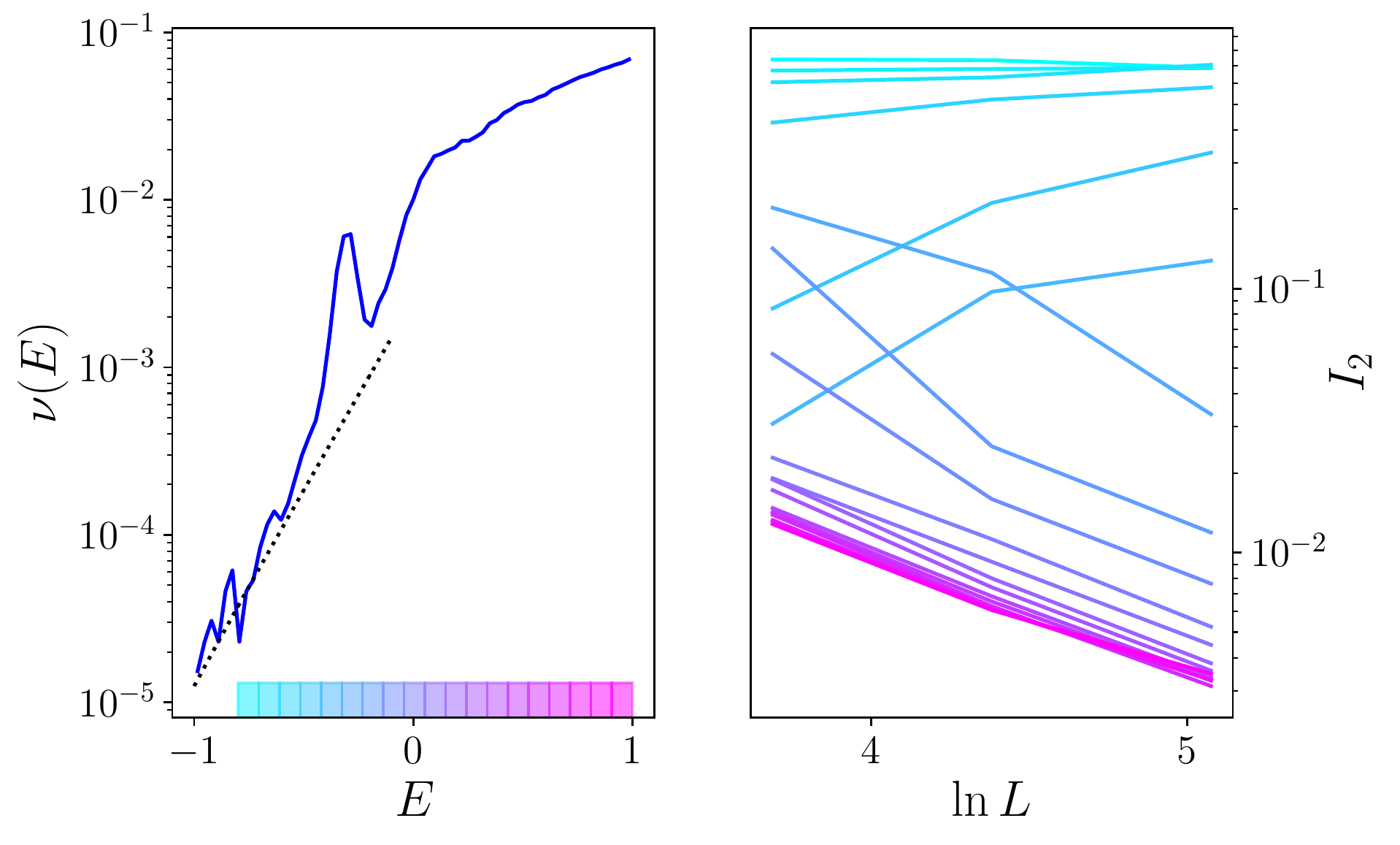}}
\caption{Density of states (left, $L=40$) and system--size dependence of the IPR (right) in 2D system for $\alpha=0.8 $
and $W=0.9$. Data in this plot are provided in terms of original energy $E$,
 to comply with notations used in Figs.~\ref{fig:scba},\ref{fig:dos1d}. Black dashed line: tail DOS, according to Eq. (\ref{DoS2}).}
\label{fig:dos2d_hi}
\end{figure}
We start the discussion of two-dimensional system by exploring the DOS and basic properties of the wavefunctions. The DOS for relatively strong disorder and $\alpha=0.8$ is shown in the Fig. \ref{fig:dos2d_hi}, left panel. There is no qualitative difference between 2d and 1d results, which were summarized above (see Figs. \ref{fig:scba} and \ref{fig:dos1d}). In particular, the shape of the tail DOS is well described by Eq. (\ref{DoS2}, which gives a result, shown by a dashed black line. It  is also instructive to explore the system size and energy dependence of the IPR, shown on the Fig. \ref{fig:dos2d_hi}, right panel (the color code of the right panel corresponds to energy bins, indicated in the bottom part of the left panel). It is clear that the tail states are localized on $O(1)$ sites (cyan curves), while the bulk states are neither localized nor ergodic (the slope of the magenta lines is substantially reduced with respect to it ergodic value, corresponding to $I_2\propto L^{-2}$). As we will show later, this is a weakly multifractal behavior, typical for 2D Anderson localization problem.

Let us discuss the localization properties of the wavefunctions in more details. At positive energies and relatively far from the  edge, one expects all eigenstates to be localized due to
quantum interference (weak localization) effects in dimensions $d=1,2$. There are two different questions
to be considered in this respect: a) dependence of typical localization length $L(E)$ as function of energy $E$,
and b) asymptotic behavior of a wavefunction modulus $\psi(r)$ at large distance from its maximum.
The answer to the latter question is rather obvious:  due to the presence of long-range hops with amplitude
$\propto -1/r^\beta$, localization in our problem is not exponential, but governed by the same power-law with
exponent $\beta = d+\alpha$.  Then localization length can be determined via IPR value
$I_2 \equiv \sum_r \psi^4(r) = [L(E)]^{-d_2}$, where $d_2$ is the fractal exponent. 
Note that power-law tail $\psi(r) \propto r^{-\beta}$ does not lead to any problem with
 convergence of the integral for $I_2$ at large distances.

In  the case of large effective
conductance $g(E) \gg 1$ in 2D, the fractal exponent is close to the space dimension, for the orthogonal symmetry class one has\cite{Wegner, FE95}
\begin{equation}
\label{eq:d2}
2-d_2 = \frac{2}{\pi g}
\end{equation}  
and fractality does not play any role: $I_2^{typ} = C_2 L^{-2}(E)$, where $C_2$ is a number somewhat larger than unity.
 We define effective dimensionless conductance $g$ similar to the usual 2D metal definition
$g = (2\pi\hbar/e^2)\sigma_\square = 2\pi\nu D$. Then localization length  follows weak-localization 
formula $L(E) = l(E)e^{\pi g(E)/2}$, where $l(E)$ is the effective mean-free-path.
Below in this Section we determine $g(E)$ and $l(E)$ behavior for $d=2$ and compare the result for $L(E)$ with
numerical data for IPR.  Note that it is more convenient to use here shifted energy argument $\epsilon$ defined 
in Eq.(\ref{Sigma2z}); below we express all energy-dependent  quantities in terms of $\epsilon$.



We will see now that at small positive (renormalized) energies effective conductance $g(\epsilon)$ of our model is large and
semi-classical calculation is adequate.  Then diffusion coefficient $D = \frac12 v^2(\epsilon)\tau(\epsilon)$, where 
$v(\epsilon) = \partial E(p)/\partial p|_{p=p(\epsilon)}$ is the group velocity and $\tau(\epsilon)$ 
is the transport scattering time.
Effective conductance  $g(\epsilon) = 2\pi \nu_0(\epsilon) D(\epsilon)= \pi v^2(\epsilon)\nu_0(\epsilon)\tau(\epsilon) $. 
 For short-range disorder we discuss here,
$1/\tau(\epsilon)$ coincides with quantum out-scattering rate $2\pi\nu_0(\epsilon)W^2$\, (here we neglect
contribution to the DOS from  strongly localized states present in Eqs.(\ref{tau-out},\ref{Sigma2wm})).
Then effective conductance is given by
\begin{equation}
g(\epsilon) = \frac{v^2(\epsilon)}{2 W^2} = c W^{-2}\epsilon^{2-2/\alpha},
\label{gE}
\end{equation}
with $c=2^{\frac{2}{\alpha}-1} \alpha^2$. The conductance $g(E)$
grows with $E$ decrease to zero in the whole range of interest, $\alpha < 1$.  To write Eq.(\ref{gE}) we used
normalization defined by Eq.(\ref{Pi00}) which is convenient here, since  results of this Section depend on low-momentum
asymptotic of $\Pi(\mathbf{q})$ only.

\begin{figure}[tbp]
\centerline{\includegraphics[width=0.75\textwidth]{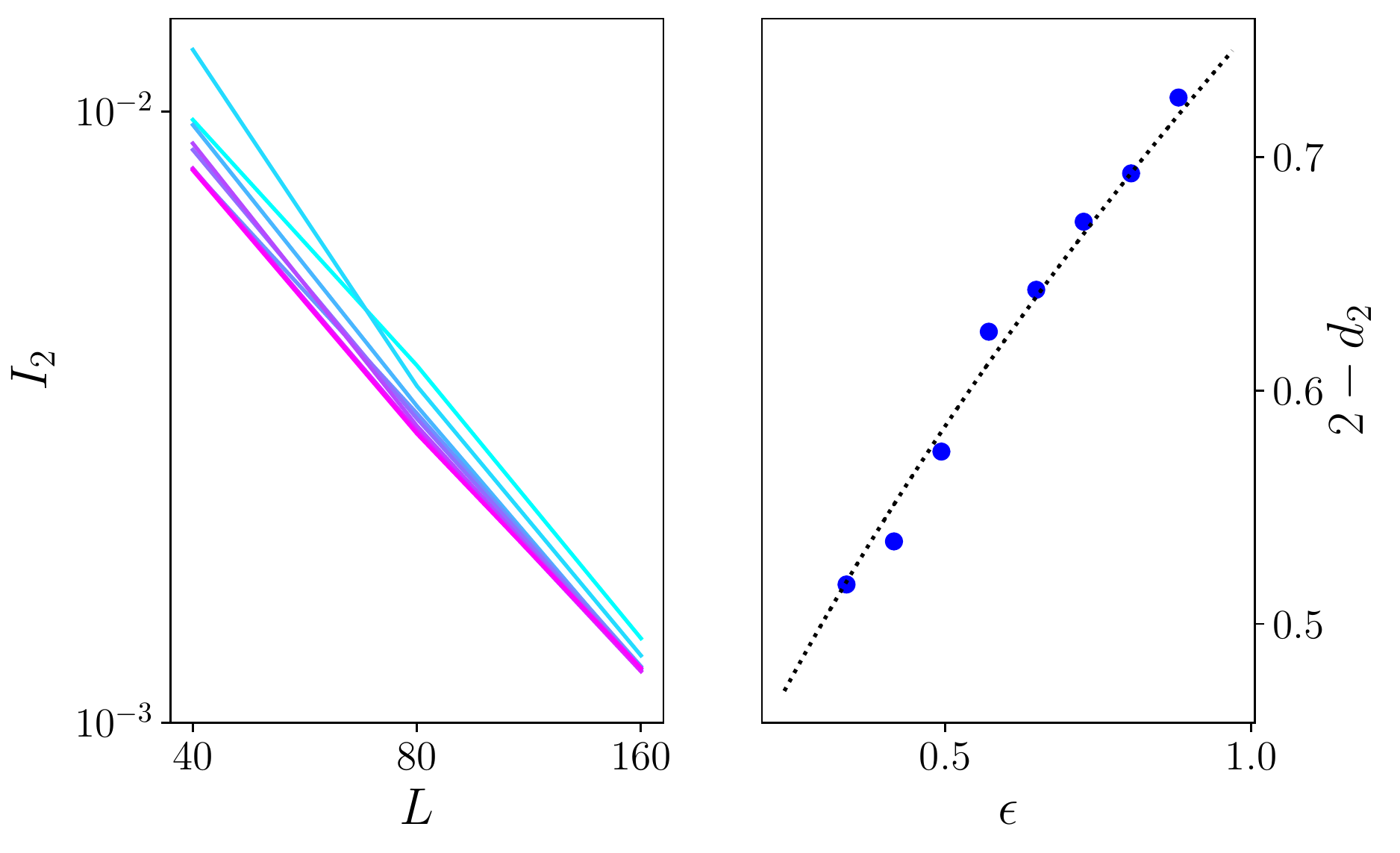}}
\caption{Left: IPR as a function of system size for $\alpha=0.8$ and $W=0.77$ for $E$ in the range [0.3, 1] (from cyan to magenta). Right: anomalous dimension $d_2$ as function of energy, as determined by IPR on system sizes $L=80; 160$. Dashed line: Eq. (\ref{gE}) together with Eq. (\ref{eq:d2}).}
\label{fig:dos2d}
\end{figure}

Mean free path $l(\epsilon)=v(\epsilon)\tau(\epsilon)$ is
strongly $\epsilon$-dependent as well:
\begin{equation}
l(\epsilon) = \frac{2^{\frac{3}{\alpha}}\alpha^2}{W^2}  \epsilon^{2-3/\alpha}
\label{lE}
\end{equation}
Due to logarithmically growing quantum interference corrections~\cite{2DWL}, all eigenstates are localized in 2D,
but localization length  $L(\epsilon)$ is very long at low $\epsilon \ll \Pi_0$ and for weak disorder $W \ll 1$:
\begin{equation}
 L(\epsilon) = \frac{2^{\frac{3}{\alpha}}\alpha^2}{W^2}  \epsilon^{2-3/\alpha}
\exp\left[
\frac{2^{\frac{2}{\alpha}} \pi \alpha^2}{4 W^2}\epsilon^{2-2/\alpha}
\right]
\label{LE}
\end{equation}
In Fig. \ref{fig:dos2d}, we compare the predictions of Eqs. (\ref{gE}) and (\ref{eq:d2}) with the results of exact diagonalization. On the left panel, we show the size-dependence of the inverse participation ratio for $\alpha=0.4$ and $W=0.22$. The curves are approximately straight lines in the log--log scale, which corresponds to the gradually (with system size) evolving power law. On the right panel, the corresponding power law (as extracted from the evolution between $L=80$ and $L=160$ system sizes) is shown as function of energy, together with theoretical prediction (dashed line, without fitting parameters).

\section{1D case at positive energies: localization length versus energy}

\begin{figure}[tbp]
\centerline{\includegraphics[width=0.75\textwidth]{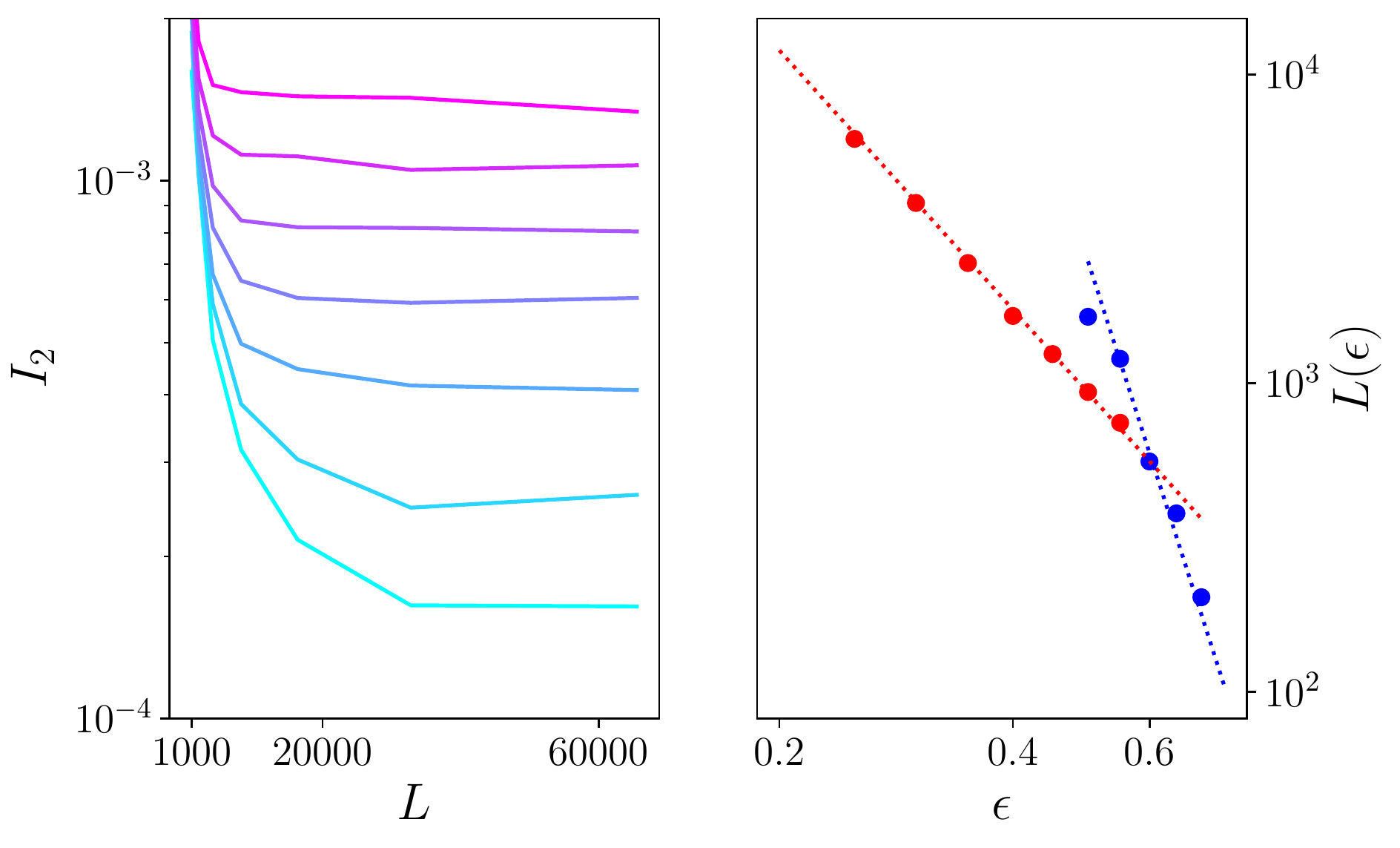}}
\caption{Left: IPR as a function of system size for $\alpha=0.4$ and $W=0.22$ for $E$ in the range [0.25, 55] (from cyan to magenta). Right: localization length as function of energy (log-log scale), red: $\alpha=0.4,\;W=0.22$, blue: $\alpha=0.2,\;W=0.32$. Dots: $L(\epsilon)=1/I_2(N=65536)$, dashed lines: $\propto \epsilon^{2-2/\alpha}$.}
\label{fig:ipr1d}
\end{figure}

Usual one-dimensional Anderson model with local disorder is known~\cite{Berezinsky} to be characterized by the
localization length $L(\epsilon)$ that coincides with backward scattering length  $l_\epsilon = \tau_\epsilon v_\epsilon$, where 
$\tau_\epsilon$ is the transport scattering time.  In our problem group velocity 
$v_\epsilon = d\Pi(q)/dq|_{q=q(\epsilon)} = 2^{1/\alpha} \alpha \epsilon^{1-1/\alpha}$, where we used again normalization condition (\ref{Pi00}).
Scattering rate $1/\tau_\epsilon = 2\pi \nu(\epsilon) W^2$, where $\nu(\epsilon) = 1/\pi v_\epsilon = (2^{1/\alpha} \pi\alpha)^{-1} \epsilon^{1/\alpha -1}$.
As a result, the estimate for the localization length $L(E)$ as well as backward scattering length $l_E$ reads 
\begin{equation}
L(\epsilon) \approx l_\epsilon = 2^{\frac{2}{\alpha}-1}\frac{\alpha^2}{W^2} \epsilon^{2-2/\alpha}
\label{LE1}
\end{equation}
The result (\ref{LE1}) is limited to the small-disorder case $W \ll 1$.
A comparison with direct numerical computation of $L(E)$ defined in terms of the IPR integral $I_2$ 
in shown in Fig.\ref{fig:ipr1d} for the case $\alpha=0.4$ and small disorder $W=0.22$, in red, and for
$\alpha=0.2$ and $W=0.32$ in blue.
 Energy-dependence of localization length perfectly
follows $\epsilon^{2- 2/\alpha}$ scaling as expected from Eq.(\ref{LE1}).

Both 2D and 1D results for the localization length, Eqs.(~\ref{LE},\ref{LE1}) show sharp increase of $L(\epsilon)$
upon decrease of $\epsilon$. On the other hand, negative-$\epsilon$ eigenstates are strongly localized (see Sec.4).
What is the mechanism which determines a crossover between so different types of $L(\epsilon)$ evolution upon
decrease  of  $\epsilon$ ?
We expect it is provided by the  nearly-localized "resonances" (Sec.4.2) those presence
was neglected in the current and previous Sections. To take them into account, we need to consider contribution
to the decay rate of the large-size (nearly delocalized) eigenstates caused by their scattering by strongly-localized resonances
(which are due to large local fluctuations of the potential). Prerequisite  for such an analysis is provided by 
Eqs.(\ref{tau-out},\ref{Sigma2wm}) in the end of Sec.4.4.

\section{Conclusions}

We have studied low-dimensional (1D and 2D)  Anderson models with local Gaussian disorder and unusual non-analitic kinetic energy
$E(q)\propto q^\alpha$  with $\alpha < d/2$.  Our main finding include the following observations:\\
 i) localized states at negative energies are due to
local (one-site) fluctuations of disorder potential and density of those states decreases as simple exponential,
see Eq.(\ref{rho-lin}), in a broad range of parameters, as long as disorder strength parameter $W$ is small. \\
ii) While typical weak fluctuations of disorder potential leads to the shift (\ref{Sigma2z}) of the apparent band edge,
the relevant energy argument relevant for the DOS in the exponential tail remains not shifted; this unusual feature is
related with special nature of fluctuations leading to the tail states, described above under item i).\\
iii) In a narrow energy range close to the renormalized band edge,
two different types of eigenstates co-exist at the same energies: delocalized and strongly localized; first of them are 
descendants of plane waves of the bare spectrum, while second are related with strong local fluctuations of disorder,
those strength is insufficient to produce localized (negative-energy) states. \\
iv) The picture described under the item iii)
exists for sub-critical magnitudes of disorder $W < W_c$,  while at $ W > W_c$ all eigenstates are localized.\\
v) At moderately small positive energies all eigenstates are localized, similar to usual 1D and 2D weak localization problem,
but localization length $L(E)$ grows sharply with $E$ decrease.


.
\end{document}